\PassOptionsToPackage{unicode}{hyperref}
\PassOptionsToPackage{hyphens}{url}
\documentclass[
  11pt,
]{article}
\usepackage{xcolor}
\usepackage{amsmath,amssymb}
\usepackage{iftex}
\ifPDFTeX
  \usepackage[T1]{fontenc}
  \usepackage[utf8]{inputenc}
  \usepackage{textcomp} 
\else 
  \usepackage{unicode-math} 
  \defaultfontfeatures{Scale=MatchLowercase}
  \defaultfontfeatures[\rmfamily]{Ligatures=TeX,Scale=1}
\fi
\usepackage{lmodern}
\ifPDFTeX\else
\fi
\IfFileExists{upquote.sty}{\usepackage{upquote}}{}
\IfFileExists{microtype.sty}{
  \usepackage[]{microtype}
  \UseMicrotypeSet[protrusion]{basicmath} 
}{}
\makeatletter
\@ifundefined{KOMAClassName}{
  \IfFileExists{parskip.sty}{%
    \usepackage{parskip}
  }{
    \setlength{\parindent}{0pt}
    \setlength{\parskip}{6pt plus 2pt minus 1pt}}
}{
  \KOMAoptions{parskip=half}}
\makeatother
\usepackage{longtable,booktabs,array}
\usepackage{calc} 
\usepackage{etoolbox}
\makeatletter
\patchcmd\longtable{\par}{\if@noskipsec\mbox{}\fi\par}{}{}
\makeatother
\IfFileExists{footnotehyper.sty}{\usepackage{footnotehyper}}{\usepackage{footnote}}
\makesavenoteenv{longtable}
\usepackage{graphicx}
\makeatletter
\newsavebox\pandoc@box
\newcommand*\pandocbounded[1]{
  \sbox\pandoc@box{#1}%
  \Gscale@div\@tempa{\textheight}{\dimexpr\ht\pandoc@box+\dp\pandoc@box\relax}%
  \Gscale@div\@tempb{\linewidth}{\wd\pandoc@box}%
  \ifdim\@tempb\p@<\@tempa\p@\let\@tempa\@tempb\fi
  \ifdim\@tempa\p@<\p@\scalebox{\@tempa}{\usebox\pandoc@box}%
  \else\usebox{\pandoc@box}%
  \fi%
}
\def\fps@figure{htbp}
\makeatother
\usepackage[authoryear,round]{natbib}
\usepackage{etoolbox}
\AtBeginEnvironment{longtable}{\footnotesize}
\usepackage[margin=1in]{geometry}
\usepackage{microtype}
\usepackage{bookmark}
\IfFileExists{xurl.sty}{\usepackage{xurl}}{} 
\hypersetup{
  pdftitle={Demand Under a Mechanical Price: Elasticity and Welfare in Ethereum's Data-Availability Market},
  pdfauthor={Farbod Ghasemlu},
  hidelinks,
  pdfcreator={LaTeX via pandoc}}

\title{Demand Under a Mechanical Price: Elasticity and Welfare in
Ethereum's Data-Availability Market}
\author{Farbod
Ghasemlu\thanks{Independent researcher. Email: \href{mailto:farbodghasemlu@outlook.com}{farbodghasemlu@outlook.com}. ORCID: \href{https://orcid.org/0009-0009-2303-5672}{0009-0009-2303-5672}.}}
\date{September 2026}

\begin{document}
\maketitle
\begin{abstract}
Ethereum sells data-availability capacity to rollups through ``blobs''
priced by a controller: the blob fee rises and falls exponentially with
cumulative usage above or below a target, and no auction sets it. We
measure what this mechanical price costs. The Pectra upgrade of 7 May
2025 doubled blob capacity on a date fixed weeks in advance, while the
market was congested, and the blob fee fell from 0.82 mETH per blob
(about \$1.40) to its 1-wei floor. Rollups' blob purchases rose 24\%.
The 95\% interval, built from the same estimator at 516 placebo dates,
runs from 8\% to 40\%, so removing the entire price raised purchases by
at most 40\%. The implied elasticity at the pre-shock price is -0.19,
and no value below -0.31 is consistent with the interval. The estimate
includes any rise in L2 activity that coincided with the fork, so it
bounds the price response from above. With demand this inelastic, the
controller works mainly as a transfer. Rollups paid \$14.8 million in
blob fees under the first blob regime, 67\% of it in two quarters when
capacity was full and any market-clearing mechanism would have collected
comparable scarcity rents. Where capacity was slack, the reserve price
introduced in December 2025 cost at most 2.6 ETH (\$8,039) in deadweight
loss under the most elastic demand the data allow. Mempool records for
5.4 million blob transactions show that when the fee sits at its floor,
capacity is slack: transactions miss about one block and are almost
never priced out.
\end{abstract}

\section{1. Introduction}\label{introduction}

Since March 2024, rollups have bought their data-availability capacity
on Ethereum at a price set by formula. The EIP-4844 blob market
\citep{eip4844} fixes a target and a maximum number of blobs per block
and moves the blob fee exponentially in the cumulative gap between usage
and target. The fee has a floor of 1 wei per unit of blob gas. In
December 2025 a second rule, EIP-7918 \citep{eip7918}, added a reserve
price tied to the execution base fee. Neither rule looks at bids.

Whether it matters that a price is mechanical depends on one number, the
price elasticity of demand. If demand is elastic, a mis-set price
changes who gets capacity and how much of it is used, and the welfare
cost is allocative. If demand is inelastic, the same mis-set price
changes mostly who pays, and the cost is a transfer. This paper
estimates that elasticity and turns it into a welfare accounting.

Estimating it is harder than the volume of on-chain data suggests. The
blob fee is common to every buyer in a block, so cross-sectional
comparisons carry no price variation, and day fixed effects absorb all
of it. Within a regime the fee is an outcome of demand by construction.
For long stretches it sits at a floor of 1 wei, where a log-price
elasticity is not economically meaningful. We therefore use a scheduled
supply shock. The Pectra upgrade \citep{eip7691} doubled the blob target
from 3 to 6 per block on 7 May 2025, a date fixed by client releases
weeks earlier, at a time when usage had been at target for months. The
fee fell from 0.82 mETH per blob to the floor within a day.

We fixed the specification in writing and committed it before estimating
it (commit \texttt{811ab9b} in the replication repository), and we
report every judgement call in a public decision log together with a
specification curve. That specification named blobs per L2 transaction
as the headline outcome. The data later showed that L2 activity itself
responded to the fork, which makes that outcome a poor measure of the
price response; we promoted the unnormalized specification, listed in
the same document, to primary and recorded the revision and its reason
alongside the original text (Section 6.2).

The paper makes five contributions.

\textbf{1. A bound on the demand elasticity for blob capacity.} When the
price of a blob fell from about \$1.40 to zero, rollups' blob purchases
rose 24\%, and the reported interval rules out increases larger than
40\%. The elasticity at the pre-shock price is -0.19, with an elastic
bound of -0.31. Blobs bought per L2 transaction did not change (-2.8\%),
so the rise came through L2 activity. A price cut of 100\% leaves little
room for the objection that the sample lacks variation.

\textbf{2. A welfare decomposition that separates transfers from
allocative loss.} Rollups paid 4,704 ETH (\$14.8 million) in blob fees
under the first regime. 67\% of this fell in 2024-Q4 and 2025-Q1, when
usage averaged 97\% of target and the binding constraint was capacity.
In slack periods the fee is either at a floor worth nothing or at a
reserve whose deadweight loss is at most 2.6 ETH over 296 days.

\textbf{3. Evidence that the floor regime is slack capacity, which
corrects a natural reading of the mechanism.} A fee pinned at its floor
looks like a price that has failed to clear the market, and one might
expect allocation by queue. We match 5.4 million blob transactions to
their first sighting in the public mempool. Transactions first seen at
the floor miss 1.03 blocks on average in the first regime, the fewest of
any state, and 0.08\% of the blocks they miss price them out.

\textbf{4. Two identification strategies that fail for mechanical
reasons.} The later capacity increases, BPO1 and BPO2 \citep{eip7892},
have no first stage: usage was already below target and the reserve set
the price. Variation in the execution base fee cannot satisfy the
exclusion restriction for the blob price, because every blob rides in a
transaction that pays execution gas at that same base fee. Reporting
both saves other researchers the attempt.

\textbf{5. Verification of the blob schedule from the protocol's own
update rule.} We recompute the excess-blob-gas state of all 6,650,443
blocks from their parents. Each regime's rule explains every block, and
the rules separate the regimes: the EIP-4844 rule mispredicts 29.5\% of
blocks after the Fusaka upgrade, although Fusaka left the blob target
and maximum unchanged.

\section{2. Related literature}\label{related-literature}

The closest economic literature studies transaction-fee mechanisms.
\citet{roughgarden2020} and \citet{roughgarden2024} analyse EIP-1559
\citep{eip1559}, the execution-gas analogue of the blob controller, as a
mechanism design problem, asking whether users, block producers and
coalitions have incentives to follow it. \citet{ferreira2021} study
dynamic posted-price mechanisms for the same market, and
\citet{leonardos2021} study the dynamics of the base-fee update rule.
Empirical work on EIP-1559 by \citet{reijsbergen2021} and
\citet{liu2022} measures fees and waiting times after its introduction.
In Bitcoin, \citet{huberman2021} model fees as the price users pay to
avoid delay when capacity is fixed, and \citet{easley2019} trace how
fees evolved as block rewards fell. Our setting differs in two respects.
The buyers number a few dozen rollups, and the product is data capacity
that rollups must buy to settle their own customers' transactions. The
question also differs: we ask what the mechanical price costs, which
requires the demand elasticity, and prior work does not estimate it for
blobs.

Work on the blob market itself is recent. \citet{crapis2023} analyse the
economics of EIP-4844 and rollups' posting strategies, including sharing
blobs, and \citet{lee2025} show with data from its first six months that
the fixed 128 KiB blob forces small rollups either to post partly empty
blobs or to post less often. \citet{mamageishvili2023} derive
batch-posting strategies for rollups facing a variable base-layer price.
\citet{park2024} describe the first months of the blob market, including
rollup transaction dynamics and the blob fee. \citet{huang2024} document
the migration of rollups from calldata to blobs and model the game
between block builders and rollups; they find that 29.5\% of
blob-carrying blocks are built in a way that earns the builder less
revenue than an available alternative. \citet{heimbach2025} study
transaction- and mempool-level data from the early blob fee market,
measure the fee loss from suboptimal block packing, and examine its
congested peak periods. \citet{ambrosia2026} trace transaction costs and
throughput on Ethereum and its L2s from January 2024 to March 2026, a
period close to ours, and find that median L2 fees fell by more than
95\% as blob capacity grew. \citet{albassam2018} set out the
data-availability problem that makes rollups buy this capacity at all.
We add a design-based elasticity estimate and a welfare accounting built
on it.

Two other strands are close. \citet{eziz2025} estimates how L2 adoption
relieved congestion on the Ethereum mainnet before Dencun, and treats
the estimates for the blob era as exploratory, and \citet{gogol2024}
study rollup performance and costs during the inscription boom and the
arrival of blobs. On the L2 side, \citet{chaliasos2025} show that rollup
fee mechanisms must price L1 data availability alongside execution, and
that mispricing it opens attacks; that pricing is the channel through
which the blob fee reaches L2 users in Section 11.

The welfare question connects to the comparison of prices and quantities
as control instruments \citep{weitzman1974}, to the empirical comparison
of posted prices and auctions \citep{einav2018}, and to the design of
dynamic fee rules for blockchain resources, where
\citet{crapis2023dynamic} derive optimal update policies and use
EIP-4844 as a case study. A parallel literature derives optimal dynamic
trading fees for automated market makers. In \citet{baggiani2025} a
venue that maximises its fee revenue raises fees when external prices
signal arbitrage and lowers them to attract noise traders; in
\citet{ghasemlu2026} a liquidity provider's growth-optimal fee rises
with volatility and is constant when volatility is. In both, the fee is
chosen to serve an objective given how traders respond. The blob fee
instead moves by a fixed rule toward a quantity target and never
consults demand, which is why the demand elasticity determines what the
rule costs. The blob controller is a hybrid: a quantity target enforced
by a price that adjusts with a lag. Our finding that demand is inelastic
places it in the region where the choice of instrument matters for
revenue and volatility much more than for allocation.

On method, our design is an event study around a protocol upgrade, as in
\citet{kapengut2023} for Ethereum's move to proof of stake. We follow
the regression-discontinuity-in-time literature on short windows and
trend flexibility \citep{hausman2018}, estimate count outcomes by
Poisson pseudo-maximum likelihood \citep{santossilva2006}, base
inference for a single event on placebo dates \citep{conley2011}, report
wild cluster bootstraps for few clusters
\citep{cameron2008, kline2012, mackinnon2017, roodman2019}, and
summarise sensitivity with a specification curve \citep{simonsohn2020}.

\section{3. Institutional setting}\label{institutional-setting}

A blob is 131,072 units of blob gas, or 128 KiB of capacity that
consensus nodes keep for about 18 days. Rollups buy blobs in type-3
transactions and post their compressed transaction data into them. We
measure quantities as blobs purchased, which is capacity bought: a
rollup pays for a whole blob whether it fills it or not.

Each block header records the excess blob gas, the cumulative usage
above target carried forward from earlier blocks. The blob fee is the
exponential of the excess divided by an update fraction, with a floor of
1 wei per unit of blob gas. In the first regime, Dencun (13 March 2024
to 7 May 2025), the target was 3 blobs and the maximum 6. Pectra raised
them to 6 and 9 \citep{eip7691}; the same upgrade repriced calldata
\citep{eip7623}. Fusaka (3 December 2025) kept 6 and 9 and introduced
EIP-7918: when the execution cost of a blob exceeds its blob fee, the
excess can no longer fall, which puts a floor under the blob fee at
roughly one sixteenth of the execution base fee per unit of blob gas.
Two blob-parameter-only upgrades followed, BPO1 (9 December 2025; 10 and
15) and BPO2 (7 January 2026; 14 and 21).

Two facts from the protocol matter for the analysis. The header-implied
fee is valid in every regime, because EIP-7918 changes only how the
excess evolves and the fee is computed from the excess as before; it
matches the fee paid within 0.1\% or 1 wei for 100.0\% of 9,287,859 blob
transactions. And the regimes are identified exactly by the update rule
(Appendix A).

\section{4. Data}\label{data}

The data combine five sources. Block headers from Dencun activation
onward, collected over public JSON-RPC, supply the fee state and blob
usage of every block. Every blob-carrying transaction comes from the
Blobscan index \citep{blobscan}; rollup labels come from Blobscan and
the blob-list registry \citep{bloblist}, released under CC0. Daily L2
activity, including transaction counts and median transaction fees,
comes from growthepie \citep{growthepie}, published under CC BY 4.0.
First mempool sightings of blob transactions for the Dencun and Pectra
regimes come from the Xatu sensor network of ethPandaOps \citep{xatu},
also CC BY 4.0. ETH/USD closes come from Binance, with Coinbase as
fallback. The data availability statement at the end describes the
deposit.

Rollup labels leave 4.3\% of blobs unattributed. The share is larger by
other measures, 10.3\% of blob transactions and 10.0\% of fees, and it
varies by regime from 0.9\% of blobs under Pectra to 8.3\% under BPO1.
Unattributed demand enters every aggregate and is excluded from
rollup-level regressions, where a residual category has no meaningful
fixed effect; including it as one more unit changes the levels estimate
from 0.191 to 0.161.

Celestia and EigenDA enter only through growthepie's daily aggregates.
Rollup-level usage of these alternative layers is not observable here
(Section 12).

\section{5. Descriptive facts}\label{descriptive-facts}

Table 1 and Figure 1 describe the market. Three features shape the
analysis.

The floor is common but not universal. The fee sat at 1 wei in 24.8\% of
Dencun blocks and 48.9\% of Pectra blocks, and in none after Fusaka.

The first regime had two phases. Through 2024-Q3 usage was mostly below
target and the fee mostly at the floor; fees for all of 2024-Q3 came to
0.000004 ETH. From October 2024 until Pectra, usage sat at 97\% of
target and the daily mean fee averaged 0.84 mETH per blob with a
standard deviation of 1.26. We split the regime by calendar quarter
only; floor incidence is an outcome and cannot define a sample.

After Fusaka the reserve set the price. Usage fell to 47\% of target
under BPO1 and 30\% under BPO2, yet the fee never touched the floor, and
the reserve governed the update in 87\% of BPO2 blocks.

\begin{longtable}[]{@{}
  >{\raggedright\arraybackslash}p{(\linewidth - 10\tabcolsep) * \real{0.3253}}
  >{\raggedright\arraybackslash}p{(\linewidth - 10\tabcolsep) * \real{0.1446}}
  >{\raggedright\arraybackslash}p{(\linewidth - 10\tabcolsep) * \real{0.1325}}
  >{\raggedright\arraybackslash}p{(\linewidth - 10\tabcolsep) * \real{0.1325}}
  >{\raggedright\arraybackslash}p{(\linewidth - 10\tabcolsep) * \real{0.1325}}
  >{\raggedright\arraybackslash}p{(\linewidth - 10\tabcolsep) * \real{0.1325}}@{}}
\caption{Blob market by fork regime. Blocks and fees from block headers
and blob transactions. The floor is 1 wei per unit of blob gas; one blob
is 131,072 units of blob gas (128 KiB of purchased capacity). USD at the
daily ETH close.}\tabularnewline
\toprule\noalign{}
\begin{minipage}[b]{\linewidth}\raggedright
\end{minipage} & \begin{minipage}[b]{\linewidth}\raggedright
dencun
\end{minipage} & \begin{minipage}[b]{\linewidth}\raggedright
pectra
\end{minipage} & \begin{minipage}[b]{\linewidth}\raggedright
fusaka
\end{minipage} & \begin{minipage}[b]{\linewidth}\raggedright
bpo1
\end{minipage} & \begin{minipage}[b]{\linewidth}\raggedright
bpo2
\end{minipage} \\
\midrule\noalign{}
\endfirsthead
\toprule\noalign{}
\begin{minipage}[b]{\linewidth}\raggedright
\end{minipage} & \begin{minipage}[b]{\linewidth}\raggedright
dencun
\end{minipage} & \begin{minipage}[b]{\linewidth}\raggedright
pectra
\end{minipage} & \begin{minipage}[b]{\linewidth}\raggedright
fusaka
\end{minipage} & \begin{minipage}[b]{\linewidth}\raggedright
bpo1
\end{minipage} & \begin{minipage}[b]{\linewidth}\raggedright
bpo2
\end{minipage} \\
\midrule\noalign{}
\endhead
\bottomrule\noalign{}
\endlastfoot
Blocks & 3,004,497 & 1,504,610 & 40,084 & 203,605 & 1,897,648 \\
Target / max (blobs) & 3 / 6 & 6 / 9 & 6 / 9 & 10 / 15 & 14 / 21 \\
Blobs per block & 2.53 & 4.56 & 5.09 & 4.66 & 4.20 \\
Usage / target & 0.84 & 0.76 & 0.85 & 0.47 & 0.30 \\
Blocks at floor & 24.8\% & 48.9\% & 0.0\% & 0.0\% & 0.0\% \\
Blocks at max & 23.0\% & 14.8\% & 17.8\% & 1.4\% & 0.3\% \\
Median fee (wei/blob gas) & 2 & 1 & 8,738,366 & 2,249,648 & 6,028,680 \\
Fees (ETH) & 4,704.4 & 7.8 & 11.9 & 0.6 & 16.3 \\
Fees (USD) & 14,813,911 & 24,393 & 37,207 & 1,790 & 35,754 \\
USD per blob & 1.9511 & 0.0036 & 0.1823 & 0.0019 & 0.0045 \\
Unattributed blobs & 4.9\% & 0.9\% & 4.0\% & 8.3\% & 6.1\% \\
\end{longtable}

\begin{figure}
\centering
\pandocbounded{\includegraphics[keepaspectratio,alt={Blob market, daily, 13 March 2024 to 28 September 2026. Top: mean blob fee over all blocks (log scale). Middle: blobs per block against target and maximum. Bottom: share of blocks at the 1-wei floor. Shaded: reserve-price regime from Fusaka.}]{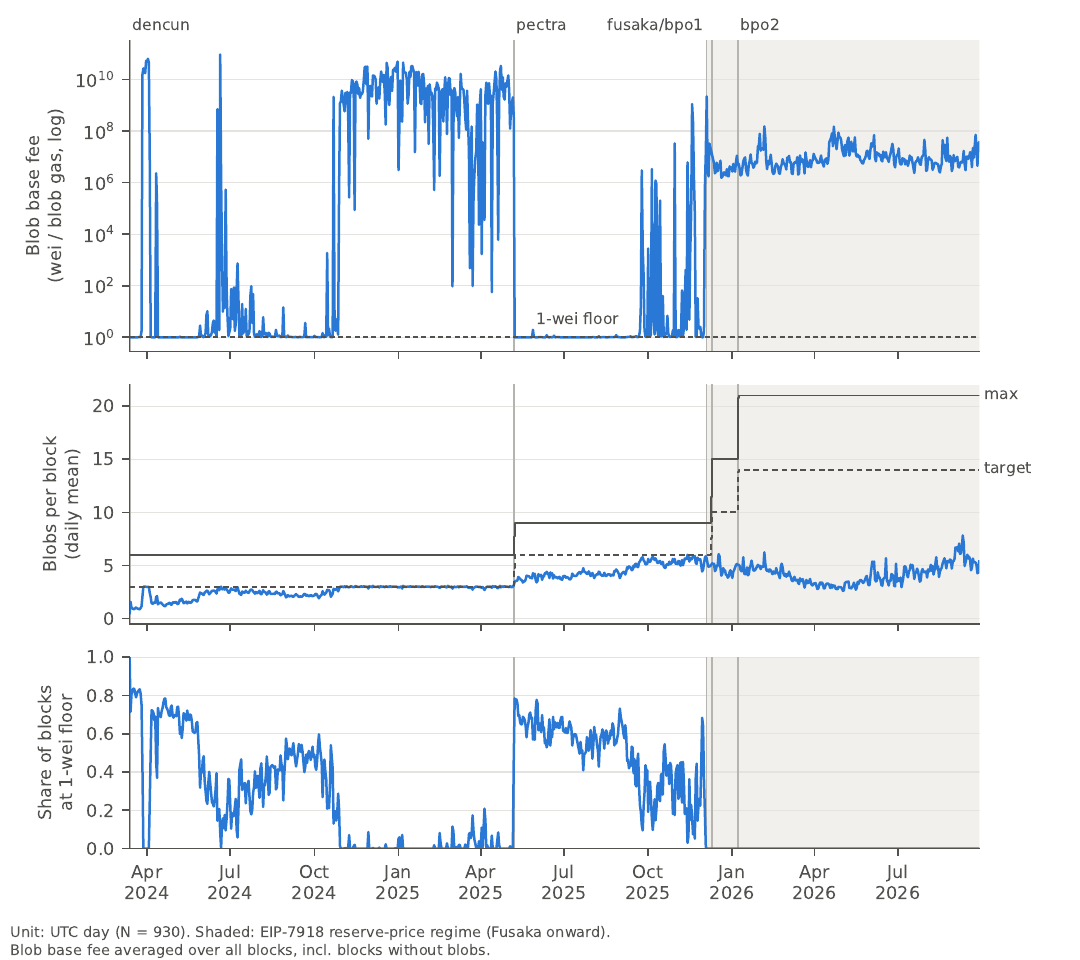}}
\caption{Blob market, daily, 13 March 2024 to 28 September 2026. Top:
mean blob fee over all blocks (log scale). Middle: blobs per block
against target and maximum. Bottom: share of blocks at the 1-wei floor.
Shaded: reserve-price regime from Fusaka.}
\end{figure}

\section{6. Identification and
estimation}\label{identification-and-estimation}

\subsection{6.1 The supply shock}\label{the-supply-shock}

Pectra is the only upgrade in the sample that moved the price
materially. In the 28 days before it, usage was at target and the mean
fee was 0.82 mETH per blob; in the 28 days after, the fee was at the
floor. The first stage, a day-level regression of the fee on a post-fork
indicator with a trend and day-of-week effects, gives a change of -0.90
mETH per blob (Newey-West SE 0.33). Blob capacity cannot be stored, so
demand on either side of the fork is not shifted by the fork itself; we
drop the fork day.

The pre-shock daily fee ranged from 0 to 4.45 mETH per blob, so the
estimates speak to prices in that range. Only 9 days of the first two
regimes had a higher daily fee, but those days carried 27\% of all fees
paid in the two regimes. Outside that range our bound does not apply.

\subsection{6.2 Estimating equations}\label{estimating-equations}

The unit of observation is a rollup-day. Let \(B_{it}\) be blobs bought
by rollup \(i\) on day \(t\), zero-filled between its first and last
post, and \(T_{it}\) its L2 transactions. We estimate by Poisson
pseudo-maximum likelihood

\[E[B_{it}] = T_{it}^{\,k} \exp(\alpha_i + \gamma_i t + \delta_{w(t)} + \beta\, \mathrm{Post}_t),\]

with rollup fixed effects \(\alpha_i\), rollup-specific linear trends
\(\gamma_i t\) and day-of-week effects \(\delta_{w(t)}\), over 28 days
either side of the fork. The primary specification sets \(k = 0\): the
outcome is blob purchases of the 15 rollups for which daily L2
transaction counts exist. We also report it for all 49 labelled rollups.

The specification we committed before estimation set \(k = 1\), entering
L2 transactions as an offset so that the outcome is blobs per L2
transaction. We chose it because ETH rose 22\% the day after Pectra,
which plausibly raised L2 activity at the moment the fee fell, and we
assumed that blob costs were too small to move L2 activity themselves.
That assumption failed (Section 11): L2 activity on blob-buying chains
rose relative to chains that do not buy blobs, so the denominator
responds to the treatment. We therefore report the per-transaction
outcome as a descriptive measure of the intensity margin and use
\(k = 0\) for inference about demand. Everything else in the
specification is as committed. The cost of the change is that the
primary estimate also contains any L2 activity raised by the ETH rally,
which biases it toward a larger response; we read it as an upper bound
on the price response.

The semi-elasticity is \(\eta = \beta/\pi\), where \(\pi\) is the first
stage, in log points per mETH per blob, and the elasticity at the
pre-shock price \(\bar p\) is \(\eta \bar p\). We measure the price in
mETH per blob. A log-price coefficient identified off a move to the
floor depends on where the floor happens to be: for the intensity
outcome the same \(\hat\beta\) gives 0.0014 with the actual 1-wei floor
and 0.0038 if the floor were \(10^6\) wei. Table 2 reports the log-price
coefficient for completeness.

\subsection{6.3 Inference with one event and few
clusters}\label{inference-with-one-event-and-few-clusters}

The design rests on one event, so standard errors clustered by rollup
describe cross-rollup noise but not the risk of a common shock on the
fork date. Our reported intervals therefore come from placebo dates
\citep{conley2011}: we apply the identical estimator at each of 516
placebo break dates between 2024-04-11 and 2025-11-04 whose windows
exclude Pectra, and invert the distribution of placebo coefficients. For
the primary specification the placebo standard deviation of
\(\hat\beta\) is 0.063, against a clustered standard error of 0.091. The
minimum detectable effect of the design, 2.8 placebo standard deviations
(80\% power against a 5\% two-sided test), is 19\%.

Clustering by rollup has a second problem. A few rollups supply most of
the blob volume: in the primary specification Base carries 78\% of the
variance of the score, and in the intensity specification 95\%. We
report the restricted wild score bootstrap of \citet{kline2012}, the
maximum likelihood analogue of the wild cluster bootstrap-t, enumerating
all 32,768 Rademacher sign vectors for 15 clusters. In a simulation with
15 equal clusters and common day shocks it rejects a true null at 7.3\%
with a nominal 5\% level.

The two procedures address different problems, and they can disagree.
The wild bootstrap corrects the bias of clustered standard errors when
clusters are few and unequal: it takes the fork date as given and draws
its uncertainty from the dispersion of rollup-level responses, which is
wide when one rollup dominates. Placebo dates address the threat that
matters most with a single treatment date, a shock common to all rollups
that happens to coincide with the fork. They measure how often the same
estimator, applied to dates with no fork, produces a coefficient as
large, and that distribution already includes the cross-rollup noise at
those dates. It assumes the placebo dates are exchangeable with the fork
date. We report the placebo interval in the abstract and in the text,
and the wild bootstrap and clustered intervals beside it in Table 2.

\section{7. Results: a bound on the
elasticity}\label{results-a-bound-on-the-elasticity}

Table 2 reports the estimates.

\begin{longtable}[]{@{}
  >{\raggedright\arraybackslash}p{(\linewidth - 6\tabcolsep) * \real{0.2949}}
  >{\raggedright\arraybackslash}p{(\linewidth - 6\tabcolsep) * \real{0.2692}}
  >{\raggedright\arraybackslash}p{(\linewidth - 6\tabcolsep) * \real{0.2115}}
  >{\raggedright\arraybackslash}p{(\linewidth - 6\tabcolsep) * \real{0.2244}}@{}}
\caption{Pectra supply shock, 28 days either side of 7 May 2025, fork
day excluded. PPML with rollup fixed effects, rollup-specific linear
trends and day-of-week effects. Column 3 includes log L2 transactions as
an offset; its denominator responds to the treatment (Section 11), so it
describes the intensity margin and is not the primary estimate. First
stage: the mean blob fee fell by 0.90 mETH per blob (Newey-West SE
0.33). Reported interval: quantiles of the same estimator at 516 placebo
break dates. Wild bootstrap: restricted score bootstrap, Rademacher
weights, all 32,768 sign vectors in columns 1 and 3, 99,999 draws in
column 2. Minimum detectable effect: 2.8 placebo standard deviations
(80\% power, 5\% test). The elasticity bound is the elastic end of the
reported interval divided by the first stage, times the pre-shock
price.}\tabularnewline
\toprule\noalign{}
\begin{minipage}[b]{\linewidth}\raggedright
\end{minipage} & \begin{minipage}[b]{\linewidth}\raggedright
(1) Blobs bought, 15 rollups (primary)
\end{minipage} & \begin{minipage}[b]{\linewidth}\raggedright
(2) Blobs bought, all rollups
\end{minipage} & \begin{minipage}[b]{\linewidth}\raggedright
(3) Blobs per L2 tx (intensity)
\end{minipage} \\
\midrule\noalign{}
\endfirsthead
\toprule\noalign{}
\begin{minipage}[b]{\linewidth}\raggedright
\end{minipage} & \begin{minipage}[b]{\linewidth}\raggedright
(1) Blobs bought, 15 rollups (primary)
\end{minipage} & \begin{minipage}[b]{\linewidth}\raggedright
(2) Blobs bought, all rollups
\end{minipage} & \begin{minipage}[b]{\linewidth}\raggedright
(3) Blobs per L2 tx (intensity)
\end{minipage} \\
\midrule\noalign{}
\endhead
\bottomrule\noalign{}
\endlastfoot
Coefficient on Post (log points) & 0.211 & 0.191 & -0.029 \\
SE, clustered by rollup & 0.091 & 0.083 & 0.033 \\
SE, two-way (rollup, day) & 0.090 & 0.080 & 0.036 \\
95\% CI, placebo quantiles (reported) & {[}0.080, 0.337{]} & {[}0.067,
0.309{]} & {[}-0.222, 0.110{]} \\
95\% CI, wild score bootstrap & {[}-0.084, 0.416{]} & {[}-0.054,
0.339{]} & {[}-0.076, 0.091{]} \\
95\% CI, clustered & {[}0.032, 0.390{]} & {[}0.029, 0.353{]} &
{[}-0.094, 0.037{]} \\
Placebo permutation p & 0.000 & 0.000 & 0.688 \\
Largest increase not ruled out & 40.1\% & 36.2\% & 11.6\% \\
Minimum detectable effect, 80\% power & 19.3\% & 18.0\% & 24.0\% \\
Semi-elasticity (per mETH/blob) & -0.235 & -0.213 & 0.032 \\
Elasticity at pre-shock price, elastic bound & -0.31 & -0.28 & -0.10 \\
Log-price coefficient (floor-dependent) & -0.0102 & -0.0092 & 0.0014 \\
N (rollup-days) & 840 & 2,744 & 840 \\
Rollups & 15 & 49 & 15 \\
\end{longtable}

Blob purchases of the 15 rollups rose 24\% when the fee fell to the
floor (column 1). The reported interval runs from 8\% to 40\%, and no
placebo date produced a coefficient as large (permutation p below 0.002;
Figure 2). Removing a price of \$1.40 per blob raised purchases by at
most 40\%. The minimum detectable effect of the design, at 80\% power,
is 19\%, so an increase of the size we estimate is within what the
design can detect. Under exponential demand the estimate implies a
semi-elasticity of -0.235 per mETH per blob and an elasticity at the
pre-shock price of -0.19; the elastic end of the interval implies -0.31.
Across all 49 labelled rollups (column 2) purchases rose 21\%.

The wild bootstrap interval for column 1, -8\% to 52\%, does not exclude
zero. For the reasons in Section 6.3 we report the placebo interval:
with one treatment date the main risk is a common shock, which the
placebo dates measure directly. The two procedures disagree about
significance and agree on magnitude. Neither interval contains an
elasticity above 1 in absolute value.

Blobs bought per L2 transaction changed by -2.8\% (column 3), with a
reported interval from -19.9\% to +11.6\%. We do not read this as a
price response, because the denominator moved with the treatment. It
says where the rise in purchases came from: rollups bought more blobs
because they carried more L2 transactions, and the capacity bought per
transaction stayed the same.

\begin{figure}
\centering
\pandocbounded{\includegraphics[keepaspectratio,alt={Coefficients at 516 placebo break dates in the first two regimes, same estimator and window, for the primary outcome (left) and the per-transaction intensity (right). The vertical line is the Pectra estimate.}]{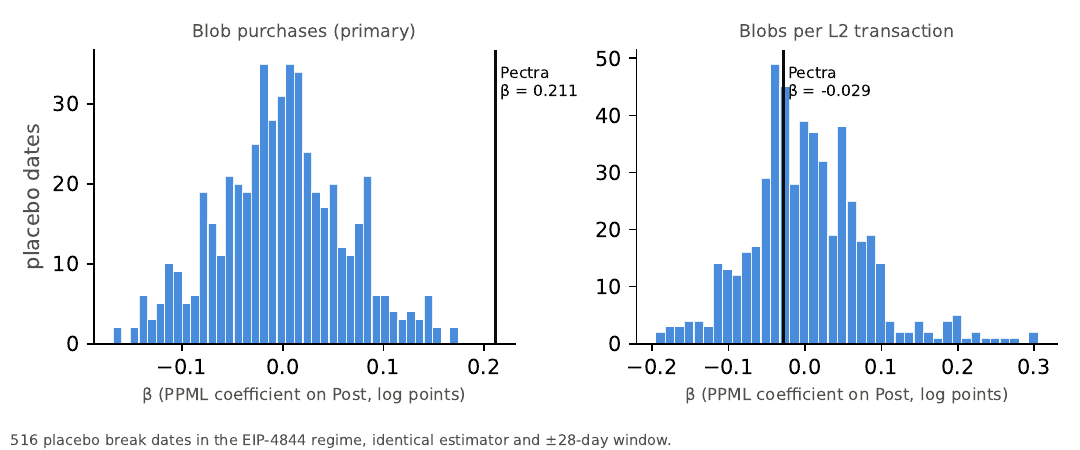}}
\caption{Coefficients at 516 placebo break dates in the first two
regimes, same estimator and window, for the primary outcome (left) and
the per-transaction intensity (right). The vertical line is the Pectra
estimate.}
\end{figure}

\textbf{Specification curve.} Across 450 specifications of the primary
outcome that vary the window, trends, the estimator, the price measure,
day-of-week effects and zero-filling, the median elasticity at the
pre-shock price is -0.22, and the 5th to 95th percentile range is -0.39
to 0.72 (Figure 3). No specification implies an elasticity above 1 in
absolute value. 13\% have the wrong sign; all of them use the 14-day
window, and none of the specifications with longer windows does. The
primary estimate is less elastic than 73\% of the curve. The
per-transaction curve (Appendix Figure 7) is centred near zero: its
median elasticity is -0.039.

\begin{figure}
\centering
\pandocbounded{\includegraphics[keepaspectratio,alt={Specification curve for the semi-elasticity of blob purchases, primary outcome. Bars: 95\% delta-method intervals from clustered errors. Lower panel: choices in each specification.}]{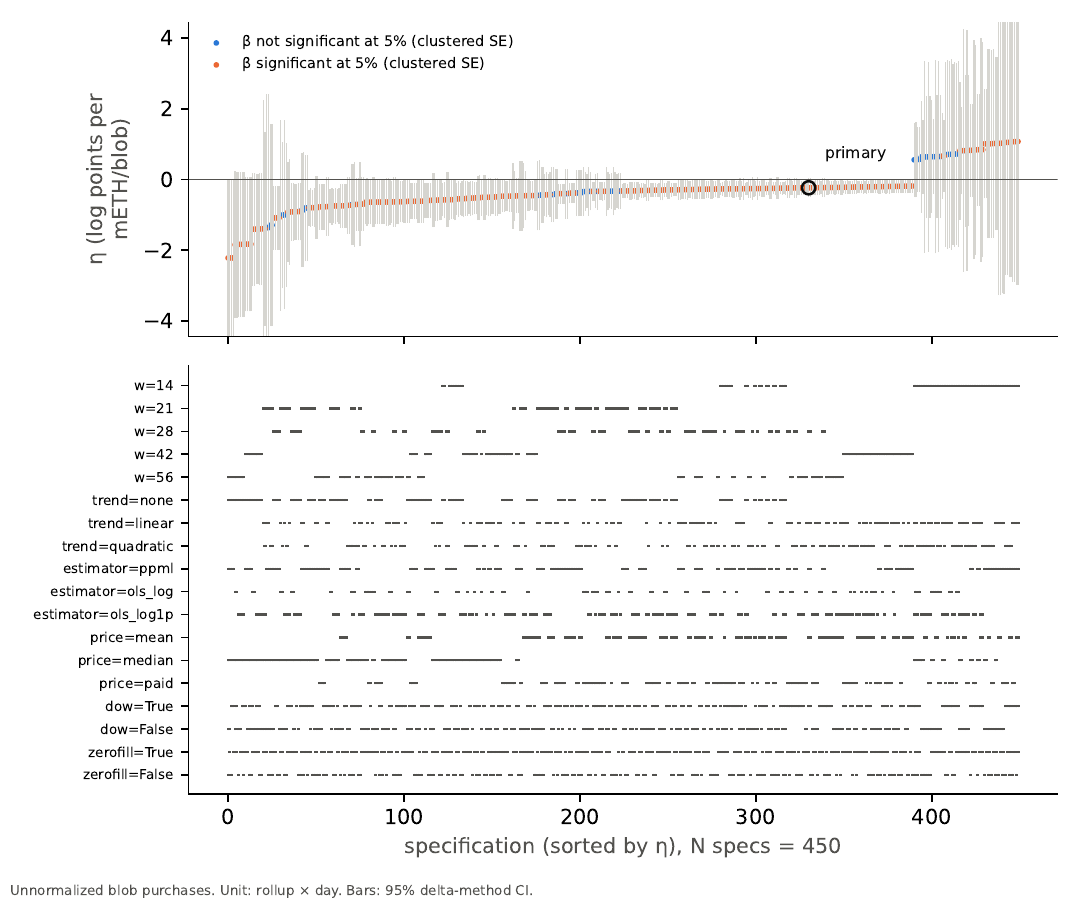}}
\caption{Specification curve for the semi-elasticity of blob purchases,
primary outcome. Bars: 95\% delta-method intervals from clustered
errors. Lower panel: choices in each specification.}
\end{figure}

\textbf{Heterogeneity.} Estimated one rollup at a time, 11 of 14 rollups
increased their purchases; the equal-weighted mean is 0.245 log points
and Base's is 0.295 (Figure 4). For blobs per L2 transaction the
per-rollup estimates split, 9 positive, with an equal-weighted mean of
0.021.

\begin{figure}
\centering
\pandocbounded{\includegraphics[keepaspectratio,alt={Coefficient on Post for blob purchases, estimated separately for each rollup, with 95\% intervals. Dashed line: pooled estimate.}]{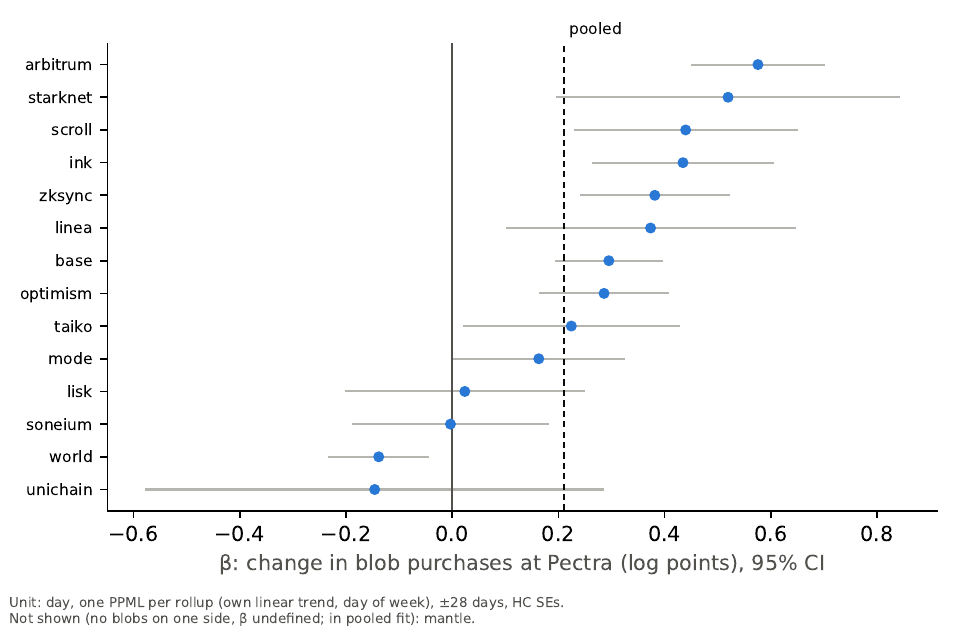}}
\caption{Coefficient on Post for blob purchases, estimated separately
for each rollup, with 95\% intervals. Dashed line: pooled estimate.}
\end{figure}

\textbf{A second source of evidence.} In the congested quarters the
daily fee had a standard deviation of 1.26 mETH per blob while usage
stayed near target. If demand shifts were as volatile as detrended L2
activity, whose standard deviation was 8.9\%, a fee clearing fixed
capacity would move by that amount divided by the semi-elasticity. The
implied semi-elasticity is 0.071 per mETH per blob, less elastic than
our primary estimate. The elastic tail of the primary curve, near -1.36,
would require daily demand shifts with a standard deviation of 171\%.
This check is a calibration, and it points toward demand at least as
inelastic as we estimate.

\textbf{Fusaka.} The reserve's introduction raised the fee from zero to
0.060 mETH per blob (about \$0.18) for the five days before BPO1; blob
purchases per L2 transaction changed by -0.011 log points (two-way SE
0.009). Five days and a price change of that size add little
information.

\section{8. The floor regime is slack
capacity}\label{the-floor-regime-is-slack-capacity}

A fee pinned at 1 wei looks like a price that has failed to clear, and a
natural reading is that allocation then happens by queue. The protocol
logic says otherwise. The fee reaches the floor only after usage has run
below target, so capacity is left over, and any rollup willing to pay 1
wei can be served. Queueing should appear instead when demand jumps
faster than the exponential fee can follow and blocks fill.

We test this with 5.4 million blob transactions matched to their first
mempool sighting, 90\% of Dencun and 92\% of Pectra blob transactions.
The 5.1 million first seen before inclusion enter the analysis; the rest
were first seen after inclusion, a sign of sensor lag. For each we count
the blocks proposed between first sighting and inclusion and classify
the state the transaction faced using only information fixed before it
could be included: whether the next block's fee was at the floor, and
whether the preceding block was full.

\begin{longtable}[]{@{}
  >{\raggedright\arraybackslash}p{(\linewidth - 12\tabcolsep) * \real{0.0621}}
  >{\raggedright\arraybackslash}p{(\linewidth - 12\tabcolsep) * \real{0.2360}}
  >{\raggedleft\arraybackslash}p{(\linewidth - 12\tabcolsep) * \real{0.0745}}
  >{\raggedleft\arraybackslash}p{(\linewidth - 12\tabcolsep) * \real{0.1366}}
  >{\raggedright\arraybackslash}p{(\linewidth - 12\tabcolsep) * \real{0.1242}}
  >{\raggedright\arraybackslash}p{(\linewidth - 12\tabcolsep) * \real{0.1677}}
  >{\raggedright\arraybackslash}p{(\linewidth - 12\tabcolsep) * \real{0.1988}}@{}}
\caption{Mempool-to-inclusion delay by market state, Dencun and Pectra,
fork days excluded. Unit: blob transaction matched to its first Xatu
mempool sighting. A missed block is one proposed after the first
sighting and before inclusion. Shares of missed blocks are pooled over
transactions that missed between 1 and 300 blocks.}\tabularnewline
\toprule\noalign{}
\begin{minipage}[b]{\linewidth}\raggedright
Regime
\end{minipage} & \begin{minipage}[b]{\linewidth}\raggedright
State when first seen
\end{minipage} & \begin{minipage}[b]{\linewidth}\raggedleft
Blob txs
\end{minipage} & \begin{minipage}[b]{\linewidth}\raggedleft
Mean blocks missed
\end{minipage} & \begin{minipage}[b]{\linewidth}\raggedright
Missed 2+ blocks
\end{minipage} & \begin{minipage}[b]{\linewidth}\raggedright
Missed blocks with room
\end{minipage} & \begin{minipage}[b]{\linewidth}\raggedright
Missed blocks pricing tx out
\end{minipage} \\
\midrule\noalign{}
\endfirsthead
\toprule\noalign{}
\begin{minipage}[b]{\linewidth}\raggedright
Regime
\end{minipage} & \begin{minipage}[b]{\linewidth}\raggedright
State when first seen
\end{minipage} & \begin{minipage}[b]{\linewidth}\raggedleft
Blob txs
\end{minipage} & \begin{minipage}[b]{\linewidth}\raggedleft
Mean blocks missed
\end{minipage} & \begin{minipage}[b]{\linewidth}\raggedright
Missed 2+ blocks
\end{minipage} & \begin{minipage}[b]{\linewidth}\raggedright
Missed blocks with room
\end{minipage} & \begin{minipage}[b]{\linewidth}\raggedright
Missed blocks pricing tx out
\end{minipage} \\
\midrule\noalign{}
\endhead
\bottomrule\noalign{}
\endlastfoot
dencun & above floor, previous block full & 681,436 & 2.25 & 24.6\% &
55.4\% & 27.20\% \\
dencun & above floor, previous block not full & 1,704,343 & 1.56 &
21.4\% & 53.0\% & 17.70\% \\
dencun & floor & 612,249 & 1.03 & 20.8\% & 62.0\% & 0.08\% \\
pectra & above floor, previous block full & 289,459 & 1.26 & 18.9\% &
47.3\% & 1.07\% \\
pectra & above floor, previous block not full & 809,019 & 0.98 & 15.4\%
& 45.8\% & 0.75\% \\
pectra & floor & 955,609 & 0.89 & 15.2\% & 54.1\% & 0.07\% \\
\end{longtable}

Transactions first seen at the floor wait the least (Table 3 and Figure
5). In Dencun they missed 1.03 blocks on average, against 2.25 when the
previous block was full, and 0.08\% of the blocks they missed priced
them out. Within rollup and day, the floor changed the probability of
missing the first eligible block by +0.018 (SE 0.006) in Dencun and
-0.006 (SE 0.004) in Pectra, against base rates of 0.42 and 0.31.

Missing the first block is common in every state. Our definition of the
first eligible block allows no time for block building, so part of the
base rate is mechanical. Block building matters for the rest: a few
builders produce most Ethereum blocks \citep{mizrach2025},
\citet{huang2024} find that builders construct many blob-carrying blocks
in ways that earn less than an available alternative, and
\citet{heimbach2025} measure fee losses from suboptimal packing of blobs
into blocks. \citet{soltani2024} model blob-transaction delay as a
queue; our measurements give its empirical counterpart and show that, at
the floor, the queue is short. Recent models of on-chain trading treat
delay and queue position as something users buy: in \citet{bergault2026}
the priority fee controls the distribution of execution delay, and in
\citet{capponi2026} paid priority and discrete block clearing determine
which traders participate. For blob transactions at the floor, the delay
we measure is about one block and the blob fee almost never decides it.
Of the delayed floor transactions, 53\% met at least one full block:
usage arrives in bursts, and a block can fill even when the running
excess is zero. The delay in those cases is about one block, 12 seconds.
Pricing out is a feature of congestion. In congested Dencun states 27\%
of missed blocks priced the transaction out.

\begin{figure}
\centering
\pandocbounded{\includegraphics[keepaspectratio,alt={Cumulative share of blob transactions by blocks missed before inclusion, by the state when first seen. Source: Xatu / ethPandaOps, CC BY 4.0.}]{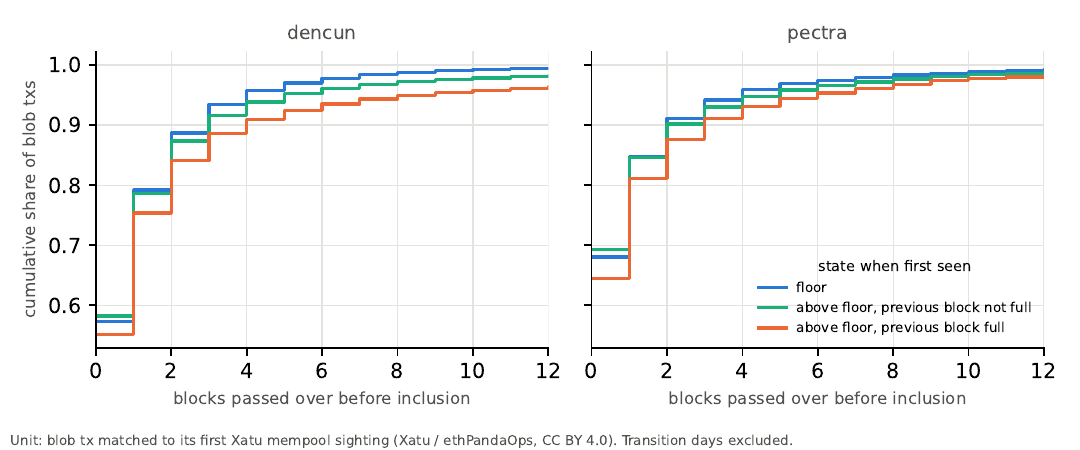}}
\caption{Cumulative share of blob transactions by blocks missed before
inclusion, by the state when first seen. Source: Xatu / ethPandaOps, CC
BY 4.0.}
\end{figure}

\section{9. Welfare: transfer and allocative
loss}\label{welfare-transfer-and-allocative-loss}

The elasticity bound decides the welfare accounting. We model demand as
exponential in the fee, \(Q(p) = A
e^{\eta p}\), calibrated each day to the observed fee and quantity.

\textbf{Slack capacity.} When usage is below target the protocol treats
the marginal cost of capacity as zero, and any positive fee excludes
some purchases worth less than the fee. At the 1-wei floor that loss is
zero to machine precision. Under the EIP-7918 reserve, when usage
averaged 32\% of target, the loss is the triangle between the demand
curve and the fee. Table 5 evaluates it at the primary estimate, at the
elastic ends of the reported intervals and at the 5th percentiles of the
specification curves. At the primary point estimate it is 0.40 ETH
(\$1,241) over 296 days, and at the elastic end of the primary interval
0.64 ETH (\$2,005). Even at the most elastic value in the table it is
2.58 ETH (\$8,039), or 10\% of the fees collected under the reserve.

\textbf{Binding capacity.} In 2024-Q4 and 2025-Q1 capacity was full and
rollups paid 3,162 ETH (\$10.4 million), 67\% of all Dencun fees. With
capacity fixed at target, some mechanism has to decide who posts, and a
uniform-price auction for the same capacity would clear where demand
meets the target. On average the controller did the same: usage was 97\%
of target. The fees were scarcity rents on fixed capacity, burned by the
protocol, and an auction would have collected rents of similar size.
What the controller adds is lag and overshooting. With demand this
inelastic, any mechanism that clears fixed capacity produces a volatile
price, and we do not separate the controller's contribution from that of
demand.

\textbf{Capacity.} Pectra cut fees from 16.9 ETH per day (\$29,182) to
effectively zero while purchases rose 24\%. The largest price episode in
the sample ended with a capacity increase, and the transfer ended with
it.

\begin{longtable}[]{@{}
  >{\raggedright\arraybackslash}p{(\linewidth - 10\tabcolsep) * \real{0.1282}}
  >{\raggedleft\arraybackslash}p{(\linewidth - 10\tabcolsep) * \real{0.1026}}
  >{\raggedleft\arraybackslash}p{(\linewidth - 10\tabcolsep) * \real{0.1410}}
  >{\raggedleft\arraybackslash}p{(\linewidth - 10\tabcolsep) * \real{0.1795}}
  >{\raggedleft\arraybackslash}p{(\linewidth - 10\tabcolsep) * \real{0.1795}}
  >{\raggedright\arraybackslash}p{(\linewidth - 10\tabcolsep) * \real{0.2692}}@{}}
\caption{Blob fees by period, with Dencun split by calendar quarter.
Fees are burned, a transfer from rollups to ETH holders.}\tabularnewline
\toprule\noalign{}
\begin{minipage}[b]{\linewidth}\raggedright
Period
\end{minipage} & \begin{minipage}[b]{\linewidth}\raggedleft
Days
\end{minipage} & \begin{minipage}[b]{\linewidth}\raggedleft
Blobs
\end{minipage} & \begin{minipage}[b]{\linewidth}\raggedleft
Fees (ETH)
\end{minipage} & \begin{minipage}[b]{\linewidth}\raggedleft
Fees (USD)
\end{minipage} & \begin{minipage}[b]{\linewidth}\raggedright
Share of all fees
\end{minipage} \\
\midrule\noalign{}
\endfirsthead
\toprule\noalign{}
\begin{minipage}[b]{\linewidth}\raggedright
Period
\end{minipage} & \begin{minipage}[b]{\linewidth}\raggedleft
Days
\end{minipage} & \begin{minipage}[b]{\linewidth}\raggedleft
Blobs
\end{minipage} & \begin{minipage}[b]{\linewidth}\raggedleft
Fees (ETH)
\end{minipage} & \begin{minipage}[b]{\linewidth}\raggedleft
Fees (USD)
\end{minipage} & \begin{minipage}[b]{\linewidth}\raggedright
Share of all fees
\end{minipage} \\
\midrule\noalign{}
\endhead
\bottomrule\noalign{}
\endlastfoot
2024-Q1 & 19 & 194,051 & 348.45 & 1,245,109 & 7.3\% \\
2024-Q2 & 91 & 1,287,299 & 671.27 & 2,293,199 & 14.2\% \\
2024-Q3 & 92 & 1,560,593 & 0 & 0 & 0.0\% \\
2024-Q4 & 92 & 1,856,822 & 1686.66 & 5,838,325 & 35.6\% \\
2025-Q1 & 90 & 1,920,236 & 1475.1 & 4,539,913 & 31.1\% \\
2025-Q2 & 37 & 773,515 & 522.92 & 897,366 & 11.0\% \\
pectra & 211 & 6,855,715 & 7.79 & 24,393 & 0.2\% \\
fusaka & 7 & 204,128 & 11.88 & 37,207 & 0.3\% \\
bpo1 & 30 & 948,791 & 0.58 & 1,790 & 0.0\% \\
bpo2 & 265 & 7,978,658 & 16.31 & 35,754 & 0.3\% \\
\end{longtable}

\begin{longtable}[]{@{}
  >{\raggedright\arraybackslash}p{(\linewidth - 8\tabcolsep) * \real{0.3665}}
  >{\raggedleft\arraybackslash}p{(\linewidth - 8\tabcolsep) * \real{0.2174}}
  >{\raggedleft\arraybackslash}p{(\linewidth - 8\tabcolsep) * \real{0.1553}}
  >{\raggedleft\arraybackslash}p{(\linewidth - 8\tabcolsep) * \real{0.1553}}
  >{\raggedright\arraybackslash}p{(\linewidth - 8\tabcolsep) * \real{0.1056}}@{}}
\caption{Deadweight loss of the EIP-7918 reserve price over 296 days
from Fusaka onward, when usage averaged 32\% of target. Exponential
demand calibrated to each day's price and quantity; the counterfactual
prices slack capacity at zero.}\tabularnewline
\toprule\noalign{}
\begin{minipage}[b]{\linewidth}\raggedright
Demand assumption
\end{minipage} & \begin{minipage}[b]{\linewidth}\raggedleft
Semi-elasticity (per mETH/blob)
\end{minipage} & \begin{minipage}[b]{\linewidth}\raggedleft
Deadweight loss (ETH)
\end{minipage} & \begin{minipage}[b]{\linewidth}\raggedleft
Deadweight loss (USD)
\end{minipage} & \begin{minipage}[b]{\linewidth}\raggedright
Share of fees
\end{minipage} \\
\midrule\noalign{}
\endfirsthead
\toprule\noalign{}
\begin{minipage}[b]{\linewidth}\raggedright
Demand assumption
\end{minipage} & \begin{minipage}[b]{\linewidth}\raggedleft
Semi-elasticity (per mETH/blob)
\end{minipage} & \begin{minipage}[b]{\linewidth}\raggedleft
Deadweight loss (ETH)
\end{minipage} & \begin{minipage}[b]{\linewidth}\raggedleft
Deadweight loss (USD)
\end{minipage} & \begin{minipage}[b]{\linewidth}\raggedright
Share of fees
\end{minipage} \\
\midrule\noalign{}
\endhead
\bottomrule\noalign{}
\endlastfoot
Primary specification, point estimate & -0.235 & 0.399 & 1,241 &
1.5\% \\
Primary specification, elastic end of reported interval & -0.375 & 0.644
& 2,005 & 2.4\% \\
Intensity specification, elastic end of reported interval & -0.122 &
0.205 & 638 & 0.8\% \\
Primary specification curve, 5th percentile & -1.363 & 2.582 & 8,039 &
9.8\% \\
Intensity specification curve, 5th percentile & -0.426 & 0.736 & 2,291 &
2.8\% \\
\end{longtable}

\section{10. Two strategies that fail}\label{two-strategies-that-fail}

\textbf{Later capacity increases.} BPO1 and BPO2 raised the target to 10
and then 14 blobs. By then usage was well below target and the fee was
set by the reserve, so extra capacity could not lower it: the mean fee
in the 28 days around BPO2 was of the order of 0.001 mETH per blob on
both sides. A capacity shock identifies a demand curve only if it moves
the price. These two did not.

\textbf{Execution base fee as an instrument.} After Fusaka the reserve
ties the blob fee to the execution base fee, and one might use spikes in
the base fee from activity unrelated to rollups, such as token launches,
as an instrument. The exclusion restriction fails before any argument
about correlated demand. Every blob is carried by a type-3 transaction
that pays execution gas at the same base fee, so the base fee raises the
cost of posting through the carrier transaction as well as through the
reserve, and both channels are proportional to it. At most, base-fee
variation identifies the response to the full cost of a post. Even that
version fails in our data. Within a day, with rollup-by-day and
rollup-by-hour-of-day fixed effects, blob purchases rise by 0.060 log
points (SE 0.020, N = 268,200 rollup-hours) for each log point of base
fee. A cost response would have the opposite sign; the estimate is what
one expects if busy hours on the base layer are also busy hours on
rollups.

\section{11. Exploratory analysis: pass-through to L2 users (post
hoc)}\label{exploratory-analysis-pass-through-to-l2-users-post-hoc}

This section was not part of the specification fixed before estimation.
We added it after the data contradicted an assumption behind that
specification, namely that blob costs are too small to affect L2 fees.
They are not, and this is the finding that led us to change the primary
outcome (Section 6.2). Before Pectra, blob cost per L2 transaction on OP
Stack chains was of the same order as the entire median L2 fee. We treat
what follows as motivation for a separate, pre-specified study.

We compare 13 rollups that buy blobs with 7 L2s and sidechains that do
not, with chain and day fixed effects over the same window (N = 1,120
chain-days). Day effects absorb the ETH rally. Blob users' median L2
fees changed by -31\% relative to the comparison chains (SE 0.20 log
points, placebo p = 0.35), and their L2 transactions rose 32\% (SE 0.11
log points, placebo p = 0.10). Within blob users, activity rose more
where fees fell more: the slope is -0.153 (SE 0.055, N = 14 chains).
Figure 6 plots the chain-level changes. Over a longer horizon,
\citet{ambrosia2026} reach the same conclusion with time-series
estimates across several upgrades. Median L2 fees fell by more than 95\%
between January 2024 and March 2026 as Ethereum's blob capacity grew,
and in a regression of the log median fee of the three largest L2s on
the blob target, from January 2025 to March 2026, each additional target
blob goes with a fall of 0.048 in log fees, about 4.6\% (their Table 7).
Because the blob target changed only at a few dates in that window,
their estimate supports the pass-through channel without separating it
from other changes over the same months.

The pattern fits demand for blobs derived from demand for L2
transactions, with the price passed through to L2 users. The evidence is
thin. It rests on one event, the activity effect clears the placebo
distribution only at the 10\% level, most comparison chains use their
own gas tokens, and OP Stack's Isthmus upgrade arrived the same week. A
second supply shock in a congested market, or rollup-level fee data at
higher frequency, would settle it. If the channel is real, it accounts
for the rise in blob purchases in Table 2 and makes blob demand derived
demand: rollups pass the price on, and their users decide how much
capacity is bought. Its implied semi-elasticity of -0.31 per mETH per
blob is less elastic than values already in Table 5, so the welfare
accounting does not change. We do not treat this as a second finding. It
is the open question with which the paper ends.

\begin{figure}
\centering
\pandocbounded{\includegraphics[keepaspectratio,alt={Change in median L2 fee and in L2 transactions, 28 days after against 28 days before Pectra, by chain. Post hoc. Source: growthepie, CC BY 4.0.}]{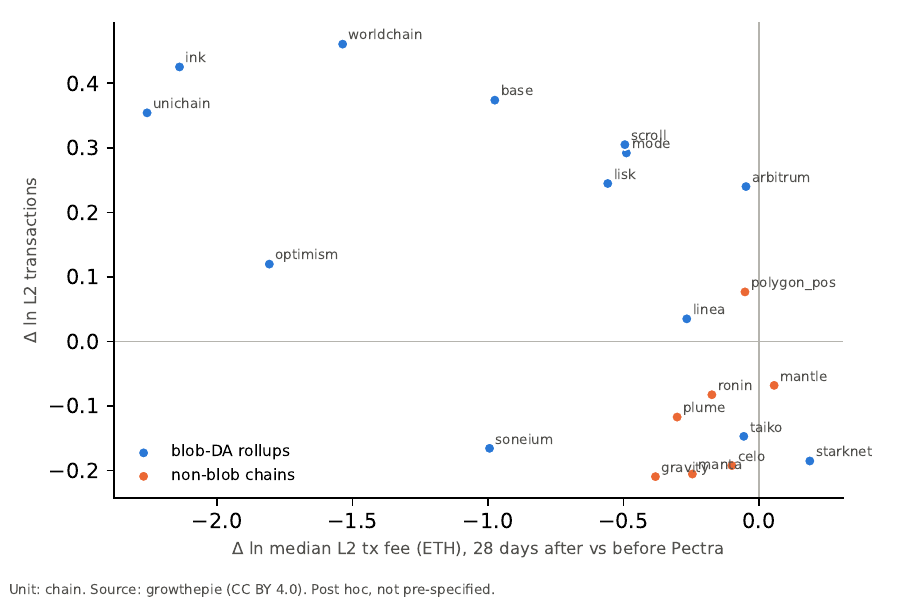}}
\caption{Change in median L2 fee and in L2 transactions, 28 days after
against 28 days before Pectra, by chain. Post hoc. Source: growthepie,
CC BY 4.0.}
\end{figure}

\section{12. Limitations}\label{limitations}

The design rests on a single event. Placebo dates measure how unusual
the estimate is, but they cannot rule out a shock specific to 7 May
2025. The ETH rally that began the next day is one such shock; it would
raise L2 activity and blob purchases, which is why we read the primary
estimate as an upper bound on the price response. Pectra also raised the
cost of calldata \citep{eip7623}. A rollup still posting calldata would
have moved to blobs for that reason, which would bias the measured
response upward; the bound is conservative in that direction.

The bound holds for fees between zero and 4.45 mETH per blob. The 9 days
with higher fees carried 27\% of the fees paid in the first two regimes,
and demand could be more elastic at those prices.

We observe capacity bought; payload is unobserved. Rollups can respond
to price by filling blobs more fully, and our outcome captures that
margin only through the number of blobs bought. For small rollups the
margin is coarse: with a fixed 128 KiB blob they choose between partly
empty blobs and less frequent posting \citep{lee2025}.

Rollup-level migration between data-availability layers is not
observable here. Celenium, the Celestia indexer that would provide
rollup-level usage of Celestia, blocks our collection host, and no
alternative source was available. The paper uses growthepie's aggregate
cross-layer series instead, which cannot show whether a given rollup
moved its data.

Unattributed blob demand is 4.3\% of blobs and 10.0\% of fees. It enters
the welfare totals and one robustness check but not the rollup-level
regressions.

\section{13. Conclusion}\label{conclusion}

Rollups' purchases of blob capacity respond weakly to its price.
Removing a price of \$1.40 per blob raised purchases by 24\%, and by no
more than 40\%, which puts the elasticity at the pre-shock price between
-0.31 and zero. With demand this inelastic, Ethereum's mechanical blob
price has worked mostly as a transfer: 4,704 ETH burned under the first
regime, most of it as scarcity rent on full capacity, against allocative
losses measured in single ETH where capacity was slack. For the design
of blob pricing, this shifts the question from allocative efficiency to
revenue, volatility and capacity.

The response we do measure came through L2 activity: blobs per L2
transaction did not change. Whether that activity rose because rollups
passed the lower price on to their users, as the post hoc evidence in
Section 11 suggests, or because of the market rally that began the next
day, is the question this design cannot settle. A pre-specified test on
the next supply shock that arrives in a congested market would.

\section{Data availability}\label{data-availability}

The data are published at Harvard Dataverse, doi:10.7910/DVN/4AOXKT,
version 1.0, under a CC BY 4.0 licence \citep{dataverse}. Exchange price
data are not redistributed; the pipeline's price script regenerates
them. The replication package, at
\url{https://github.com/farbodghasemlu/ethereum-blob-demand},
regenerates every number, table and figure from the deposit with one
command. Blob transaction data are indexed by Blobscan \citep{blobscan},
with batcher labels from blob-list (CC0) \citep{bloblist}. L2
fundamentals are from growthepie (CC BY 4.0) \citep{growthepie}. Mempool
observations are from Xatu / ethPandaOps (CC BY 4.0) \citep{xatu}. The
specification fixed before estimation (commit \texttt{811ab9b}), the
decision log and the specification curve are in the replication
repository.

\section{Appendix A. Verifying the blob
schedule}\label{appendix-a.-verifying-the-blob-schedule}

We recompute the excess blob gas of every block from its parent's
excess, usage and base fee, applying the EIP-4844 rule through Pectra
and the EIP-7918 rule from Fusaka. The parent's blob fee is priced with
the child block's update fraction; the BPO1 activation block is
consistent only with this convention. Table 6 reports the results for
6,650,443 blocks.

\begin{longtable}[]{@{}
  >{\raggedright\arraybackslash}p{(\linewidth - 10\tabcolsep) * \real{0.0645}}
  >{\raggedleft\arraybackslash}p{(\linewidth - 10\tabcolsep) * \real{0.1161}}
  >{\raggedright\arraybackslash}p{(\linewidth - 10\tabcolsep) * \real{0.1613}}
  >{\raggedright\arraybackslash}p{(\linewidth - 10\tabcolsep) * \real{0.2129}}
  >{\raggedright\arraybackslash}p{(\linewidth - 10\tabcolsep) * \real{0.2129}}
  >{\raggedright\arraybackslash}p{(\linewidth - 10\tabcolsep) * \real{0.2323}}@{}}
\caption{Excess blob gas recomputed for every block from its parent. Own
rule: EIP-4844 for Dencun and Pectra, EIP-7918 from Fusaka. The parent's
blob fee is priced with the child block's update fraction. Comparisons
within 3 wei are not scored.}\tabularnewline
\toprule\noalign{}
\begin{minipage}[b]{\linewidth}\raggedright
Regime
\end{minipage} & \begin{minipage}[b]{\linewidth}\raggedleft
Blocks checked
\end{minipage} & \begin{minipage}[b]{\linewidth}\raggedright
Unexplained, own rule
\end{minipage} & \begin{minipage}[b]{\linewidth}\raggedright
Mispredicted by EIP-4844 rule
\end{minipage} & \begin{minipage}[b]{\linewidth}\raggedright
Mispredicted by EIP-7918 rule
\end{minipage} & \begin{minipage}[b]{\linewidth}\raggedright
Reserve comparisons within 3 wei
\end{minipage} \\
\midrule\noalign{}
\endfirsthead
\toprule\noalign{}
\begin{minipage}[b]{\linewidth}\raggedright
Regime
\end{minipage} & \begin{minipage}[b]{\linewidth}\raggedleft
Blocks checked
\end{minipage} & \begin{minipage}[b]{\linewidth}\raggedright
Unexplained, own rule
\end{minipage} & \begin{minipage}[b]{\linewidth}\raggedright
Mispredicted by EIP-4844 rule
\end{minipage} & \begin{minipage}[b]{\linewidth}\raggedright
Mispredicted by EIP-7918 rule
\end{minipage} & \begin{minipage}[b]{\linewidth}\raggedright
Reserve comparisons within 3 wei
\end{minipage} \\
\midrule\noalign{}
\endhead
\bottomrule\noalign{}
\endlastfoot
dencun & 3,004,496 & 0.0000\% & 0.0\% & 37.6\% & 0.0000\% \\
pectra & 1,504,610 & 0.0000\% & 0.0\% & 46.1\% & 0.0000\% \\
fusaka & 40,084 & 0.0000\% & 29.5\% & 0.0\% & 0.0000\% \\
bpo1 & 203,605 & 0.0000\% & 76.9\% & 0.0\% & 0.0005\% \\
bpo2 & 1,897,648 & 0.0000\% & 87.0\% & 0.0\% & 0.0002\% \\
\end{longtable}

\begin{figure}
\centering
\pandocbounded{\includegraphics[keepaspectratio,alt={Specification curve for the semi-elasticity of blobs per L2 transaction (intensity margin). Bars: 95\% delta-method intervals from clustered errors. Lower panel: choices in each specification.}]{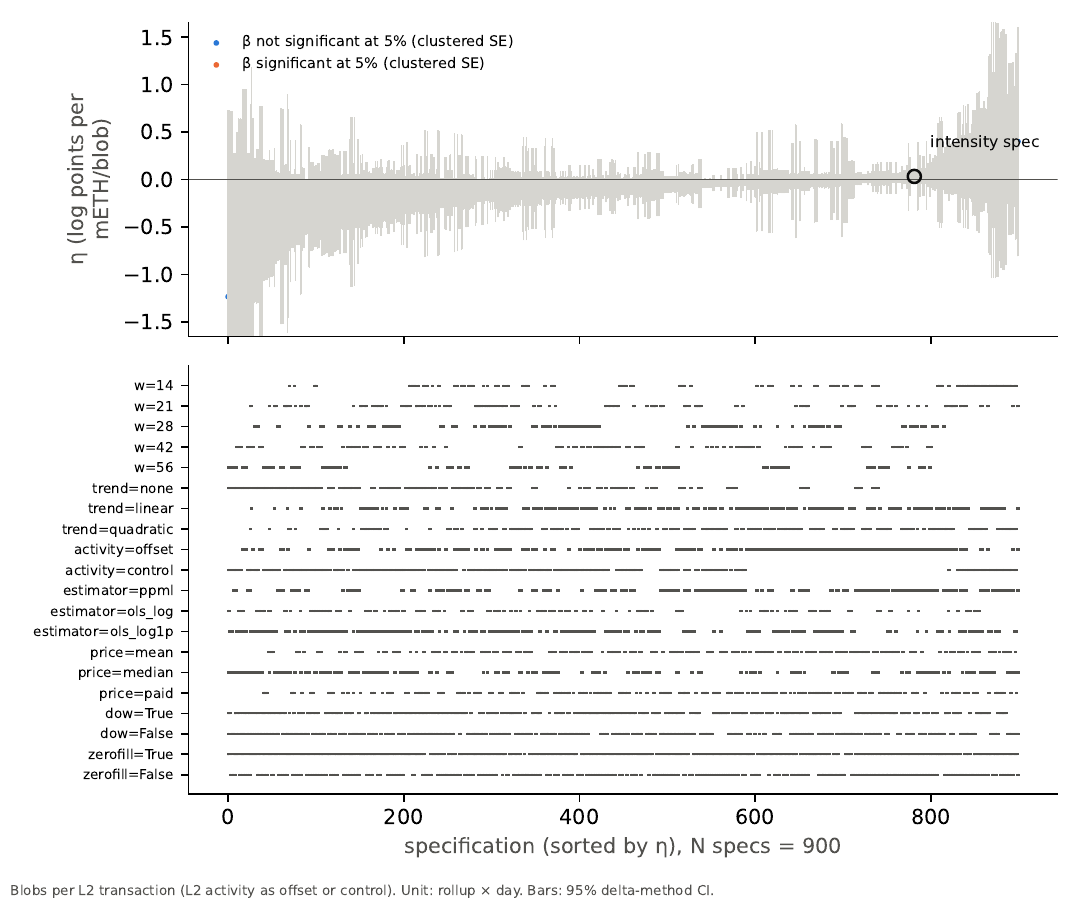}}
\caption{Specification curve for the semi-elasticity of blobs per L2
transaction (intensity margin). Bars: 95\% delta-method intervals from
clustered errors. Lower panel: choices in each specification.}
\end{figure}

Each regime's own rule explains every block. Five reserve comparisons
fall within 3 wei of equality and are not scored. The cross-rule columns
show that the rules are distinguishable: the EIP-4844 rule mispredicts
29.5\% of Fusaka blocks, and the EIP-7918 rule would have mispredicted
37.6\% of Dencun and 46.1\% of Pectra blocks. The reserve governed the
update in 29.5\% of Fusaka, 76.9\% of BPO1 and 87.0\% of BPO2 blocks.

\bibliography{references.bib}

\end{document}